# Symbolic Reasoning for Automatic Signal Placement (Extended Version)

Kostas Ferles
The University of Texas at Austin
Austin, TX, USA
kferles@cs.utexas.edu

Jacob Van Geffen
The University of Texas at Austin
Austin, TX, USA
jsv@cs.utexas.edu

Isil Dillig
The University of Texas at Austin
Austin, TX, USA
isil@cs.utexas.edu

Yannis Smaragdakis
University of Athens
Athens, Greece
smaragd@di.uoa.gr

**Abstract**
Explicit signaling between threads is a perennial cause of bugs in concurrent programs. While there are several runtime techniques to *automatically* notify threads upon the availability of some shared resource, such techniques are not widely-adopted due to their run-time overhead. This paper proposes a new solution based on static analysis for automatically generating a performant *explicit-signal* program from its corresponding *implicit-signal* implementation. The key idea is to generate verification conditions that allow us to minimize the number of required signals and unnecessary context switches, while guaranteeing semantic equivalence between the source and target programs. We have implemented our method in a tool called Expresso and evaluate it on challenging benchmarks from prior papers and open-source software. Expresso-generated code significantly outperforms past automatic signaling mechanisms (avg. 1.56x speedup) and closely matches the performance of hand-optimized explicit-signal code.

## 1 Introduction

A common challenge in concurrent programming is to coordinate access to shared resources and achieve correct synchronization between different threads. While there are many different language constructs that can be used to perform synchronization, a widely-established programming pattern is to encapsulate inter-thread coordination using *monitors* [23, 28, 39]. At a high level, a monitor encapsulates all shared state between threads and guarantees mutual exclusion. In addition, monitors perform synchronization between threads by blocking and unblocking them depending on the availability of some shared resource.

Broadly speaking, monitors can be classfied into two categories, depending on the burden they impose on the system vs. the programmer [7]. In particular, *explicit-signal monitors* typically employ *condition variables* to perform synchronization between threads and use an explicit "signal" construct to notify other threads when some shared resource becomes available. In contrast, implicit-signal (automatic) monitors provide a waituntil(P) construct such that any thread executing this statement blocks until predicate P becomes true. In implicit-signal monitors, there is no explicit signal construct, and it is the responsibility of the system to notify threads that are currently blocked on a predicate. To give the reader some intuition, Figure 1 shows the implementation of an implicit-signal monitor for the well-known readers-writers problem, and Figure 2 shows its corresponding implementation as an explicit-signal monitor.

As illustrated by the example from Figures 1 and 2, programming with implicit monitors is considerably easier because the programmer does not need to reason about when and which threads should be notified. In fact, it is well-known that many concurrency bugs are caused by erroneous signal placement in explicit-signal implementations [22, 31]. However, despite their easier programmability, implicit-signal monitors are not widely-used due to performance considerations. In particular, because the system needs to notify threads that are blocked on a predicate, run-time support for implicit-signal monitors may result in considerable overhead. For example, according to Buhr et al., automatic monitors can be 10-50 times slower than explicit signals [7]. Even though recent work by Hung and Garg proposes a more efficient implementation of automatic monitors [30], explicit-signal monitors still remain the de-facto synchronization mechanism in real-world concurrent programs.

In this paper, we propose a new solution —based on static analysis— to programming with implicit-signal monitors. Given the implementation of an implicit-signal monitor, our method automatically synthesizes an efficient and semantically equivalent explicit-signal implementation. We believe this approach has two advantages compared to prior run-time techniques: First, because our method does not require additional run-time book-keeping, it has the potential to be as efficient as a performant hand-written explicit-signal implementation. Second, because the code generated by our system can be inspected and further refined by the programmer, it is more transparent compared to automatic-signaling systems that provide run-time instrumentation.

While it is straightforward to generate *any* semantically equivalent explicit-signal implementation of an automatic-signal monitor, a key consideration is the *efficiency* of the synthesized code. In particular, the synthesized code should *not* spuriously wake up threads that are blocked on a predicate that evaluates to false.[1] In practice, this means that the generated code should not notify threads blocked on a predicate $P$, if $P$ is guaranteed to be false at the time of notification. Furthermore, whenever possible, the generated code should notify a single –rather than *all*– threads blocked on a predicate in order to avoid unnecessary context switches.

In addition to avoiding spurious wake-ups, another important efficiency consideration is to minimize the use of *conditional signals*, which notify other threads only if some condition evaluates to true. Because conditional signals require evaluating the truth value of (potentially complex) predicates at run-time, it is desirable to use unconditional signals whenever possible. In fact, while some run-time solutions, such as AutoSynch [30] avoid spurious wake-ups altogether, they may still incur significant overhead due to the frequent evaluation of predicates at run-time.

The solution that we adopt in this paper tries to minimize both the use of spurious wake-ups as well as conditional signals by performing precise static analysis of the monitor code. In particular, our method automatically generates Hoare triples, that, if valid, allow us to establish that a program fragment does *not* need to signal other threads waiting on a predicate. For program fragments where signaling may be necessary, our method generates additional Hoare triples whose validity allows us to minimize the use of conditional signals as well as *broadcast* operations that notify all threads.

In order to successfully discharge the generated Hoare triples, our method uses so-called *monitor invariants*, which are assertions that hold every time a thread enters or leaves the monitor. Our approach automatically infers these monitor invariants by combining abductive reasoning and predicate abstraction, allowing the synthesis of non-trivial invariants that involve disjunctions. Monitor invariants allow us to discharge verification conditions that could not be proven otherwise (e.g., by strengthening the precondition of the generated Hoare triples) and are therefore crucial for generating efficient explicit-signaling code.

We have implemented our proposed ideas in a tool called Expresso and evaluate the efficiency of the code generated by Expresso by comparing it against manually written explicit-signal monitors as well as the state-of-the-art

---

[1] While a thread that is spuriously woken up will "go back to sleep", this introduces significant overhead due to an unnecessary context-switch.

```
1    class RWLock {
2      unsigned int readers = 0;
3      boolean writerIn = false;
4
5      atomic void enterReader() {
6        waituntil(!writerIn);
7        readers++;
8      }
9      atomic void exitReader() {
10       if(readers > 0) readers--;
11     }
12     atomic void enterWriter() {
13       waituntil(readers == 0 && !writerIn);
14       writerIn = true;
15     }
16     atomic void exitWriter() {
17       writerIn = false;
18     } }
```

**Figure 1.** Implicit-signal monitor for readers-writers lock.

AutoSynch tool that provides run-time support for implicit-signal monitors. Our evaluation shows that the performance of the code synthesized by Expresso is an average of 1.56x faster than AutoSynch and comparable to that of hand-written code.

In all, this paper makes the following key contributions:

- We propose a novel technique, based on static analysis, for generating efficient explicit-signal implementations of implicit-signal monitors.
- We show how the automatic signal placement problem can be reduced to proving the validity of certain kinds of Hoare triples in concurrent programs.
- We introduce the notion of *monitor invariants* and show how to automatically infer them using abductive reasoning and monomial predicate abstraction.
- We implement the proposed techniques in a tool called Expresso and evaluate it by comparing against AutoSynch, a state-of-the art runtime system for implicit-signal monitors, as well as hand-written code.

## 2 Overview of Technique

In this section, we give a high-level overview of our approach with the aid of the reference Readers-Writers example, shown in Figure 1. In particular, we explain the reasoning performed by Expresso to automatically generate the code shown in Figure 2 by analyzing the implicit-signal monitor of Figure 1.

Expresso starts its analysis by inferring a *monitor invariant*, which is an assertion that holds every time a thread enters or exits the monitor. For the code in Figure 1, Expresso successfully infers the invariant $readers \geq 0$. Then, Expresso uses this invariant to determine for each conditional critical section in Figure 1 (a) if signaling is necessary, (b) whether to signal or broadcast, and (c) whether to do so conditionally or unconditionally.

***EnterReader.*** Consider a reader thread $t_r$ executing the method enterReader. To generate explicit signaling code,

we need to determine whether $t_r$ needs to notify any writer threads blocked on predicate $P_w = (\textit{readers} = 0 \land \neg \textit{writerIn})$ at line 13. Towards this goal, we ask the following question: "Assuming that a writer thread $t_w$ is blocked on $P_w$, is it possible that $P_w$ becomes true after $t_r$ executes the code in enterReader?". If the answer to this question is "no", we have established that $t_r$ does not need to signal. Thus, to prove that no signals are necessary, Expresso generates and checks the validity of the following Hoare triple:

$$\{\textit{readers} \geq 0 \land \neg \textit{writerIn} \land \neg P_w\} \text{ readers++ } \{\neg P_w\}$$

Here, the precondition states that (a) the monitor invariant holds when $t_r$ enters the monitor, (b) !writerIn must hold if $t_r$ executes readers++, and (c) $P_w$ is false, meaning that some writer thread may be blocked at line 13. The post-condition says that $P_w$ continues to stay false after $t_r$ exits the monitor. Since this Hoare triple is indeed valid, Expresso establishes that no signaling is necessary. Observe that dropping the conjunct $\textit{readers} \geq 0$ from the precondition would result in a Hoare triple that is *not* valid; thus, the monitor invariant is crucial for avoiding the signal operation in this example.

***ExitReader.*** For the exitReader method, Expresso needs to determine whether we should signal any reader threads blocked at line 6 or writer threads blocked at line 13. Using similar reasoning as in enterReader, it is easy to establish that we do not need to signal reader threads because readers-- does not affect the truth value of the predicate writerIn. Now, to determine the necessity of signaling writer threads, Expresso generates the following Hoare triple:

$$\{\textit{readers} \geq 0 \land \neg P_w\} \text{ if(readers > 0) readers-- } \{\neg P_w\}$$

Since this Hoare triple is not valid, signalling is necessary.

Next, Expresso tries to determine whether it suffices to notify a *single* writer thread or we need to notify all writers blocked at line 13. To answer this question, we ask "Is it possible that $P_w$ stays true after some writer thread $t_w$ executes enterWriter?". If not, we have proven that it is unnecessary (and wasteful) to wake up multiple threads, since $P_w$ becomes false after the first writer thread executes. Thus, Expresso generates and checks the following Hoare triple:

$$\{\textit{readers} \geq 0 \land P_w\} \text{ writerIn = true } \{\neg P_w\}$$

Since this triple is valid, Expresso has determined that broadcasting is *not* necessary.

Finally, Expresso checks whether it can signal unconditionally, meaning that $P_w$ is guaranteed to hold after the reader thread $t_r$ exits the monitor. Towards this goal, we perform the following check:

$$\{\textit{readers} \geq 0 \land \neg P_w\} \text{ if(readers > 0) readers-- } \{P_w\}$$

This Hoare triple is not valid, so Expresso signals conditionally in order to avoid a spurious wake-up.

```
1    class RWLock {
2      unsigned int readers = 0;
3      boolean writerIn = false;
4      Lock l = new ReentrantLock();
5      Condition readers = l.newCondition(),
6                writers = l.newCondition();
7      void enterReader() {
8        l.lock();
9        while(writerIn) readers.await();
10       readers++;
11       l.unlock();
12     }
13     void exitReader() {
14       l.lock();
15       if (readers > 0) readers--;
16       if (readers == 0) writers.signal();
17       l.unlock();
18     }
19     void enterWriter() {
20       l.lock();
21       while(readers != 0 || writerIn) writers.await();
22       writerIn = true;
23       l.unlock();
24     }
25     void exitWriter() {
26       l.lock();
27       writerIn = false;
28       if (readers == 0) writers.signal();
29       readers.signalAll();
30       l.unlock();
31   } }
```

**Figure 2.** Explicit-signal monitor for readers-writers lock.

***EnterWriter.*** Using similar reasoning as in enterReader, Expresso can establish that enterWriter does not need to signal any readers because the following Hoare triple is valid:

$$\{\textit{readers} \geq 0 \land P_w \land \textit{writerIn}\} \text{ writerIn = true } \{\textit{writerIn}\}$$

***ExitWriter.*** For exitWriter, Expresso establishes that it is necessary to notify both reader and writer threads. Using similar reasoning as in exitReader, we can prove that broadcasting for all *writer* threads is not necessary, however, we need to perform conditional signaling to avoid spurious wake-ups. For *reader* threads, Expresso determines that broadcasting is necessary since !writerIn continues to hold after executing the statement readers++. Furthermore, since the Hoare triple

$$\{\textit{readers} \geq 0 \land \textit{writerIn}\} \text{ writerIn = false } \{\neg \textit{writerIn}\}$$

is valid, Expresso can establish that !writerIn must be true after the writer thread exits. Thus, Expresso instruments the code to signal reader threads unconditionally.

***Summary.*** For this example, the code generated by Expresso is precisely the same one as the human-written explicit-signal implementation shown in Figure 2. Observe that Expresso can prove the gratuitousness of broadcasts, and it can also establish that enterReader and enterWriter do not need to signal. Finally, note that some of the Hoare triples generated by Expresso could not be established without the useful monitor invariant $\textit{readers} \geq 0$.

## 3 Source and Target Languages

In this section, we present some preliminary concepts related to concurrent programming and describe the source and target languages that can be used to implement implicit- and explicit-signal monitors respectively. The goal of the two languages presented here is to provide a unified theoretical framework suitable for automatic reasoning. In Section 6, we discuss how the target language can be instantiated in a concrete monitor implementation.

### 3.1 Preliminaries

In this paper, we consider a shared-memory concurrency model in which all accesses to shared resources occur inside a *monitor*. In other words, all variables accessed outside the monitor are assumed to be thread-local. We represent threads using integer identifiers drawn from the set $T \subseteq \mathbb{N}$. Because we do not impose any restrictions on the number of threads that can execute monitor code, our approach is applicable to parametrized concurrent programs.

We partition program variables used in the monitor into two disjoint sets, namely $L$ and $G$, representing thread-local and shared (global) variables respectively. As stated by the definition below, the state $\sigma$ of a monitor identifies the values of program variables for each thread.

**Definition 3.1. (Monitor state)** A monitor state, $\sigma : T \times (L \cup G) \to \mathbb{N}$, is a mapping from (thread identifier, monitor variable) pairs to a value. We require monitor states to agree on the values of shared variables for all threads; i.e.,

$$\forall t_1, t_2 \in T, v \in G.\ \sigma(t_1, v) = \sigma(t_2, v)$$

### 3.2 Source Language

Because our approach transforms an automatic-signal monitor to an explicit-signal one, we first present the source language in which automatic-signal monitors are implemented.

The syntax of our source language is presented in Figure 3. Since our implementation targets Java programs, we consider implicit-signal monitors written in a simple object-oriented language with Java-like syntax. In particular, an automatic-signal monitor consists of a set of field declarations and a set of *atomic* methods – i.e., the body of a method $m$ executes without interruption unless the thread blocks on some waituntil statement whose corresponding predicate evaluates to false. To simplify presentation, we will assume that *local variables of different methods have unique names*.

The body of each monitor method is a sequence of statements of the form waituntil($p$){$s$}, where $p$ is a predicate and $s$ is a statement (assignment, store, sequence, loop etc.). Observe that a statement $s$ is a special case of a waituntil statement whose corresponding predicate is *true*. We refer to predicate $p$ as the *guard* of the waituntil construct and to statement $s$ as its *body* and sometimes write $w = (p, s)$ to denote a waituntil statement with guard $p$ and body $s$. Given a monitor $M$, we use the notation $CCRs(M)$ to represent the set of all waituntil statements used in any method in $M$.

```
Monitor M   ::= monitor M { (fld | m)* }
Field fld   ::= f : τ
Method m    ::= atomic m(v⃗) {w*; return v; }
WUntil w    ::= waituntil(p){ s }
Statement s ::= skip | s₁; s₂ | v := e | v.f := e
              | if (p) {s₁} else {s₂}
              | while (p) {s}
```

**Figure 3.** Implicit-signal monitor language. Here, $e$ and $p$ denote expressions and predicates respectively.

While waituntil statements can only appear as top-level statements in our source language, we note that this design decision does not sacrifice expressiveness. For example, consider the code snippet `if (c) waituntil(p)`, which is not supported by our source language. Observe that the check `if(c)` can be moved outside of the monitor if `c` is not on shared data. On the other hand, if `c` does involve shared data, the condition is either checked in a logically racy way [2] or the programmer knows that `c` cannot change while the thread is blocked on `p`. In either case, the program's logic is preserved or enhanced if condition `c` is moved inside the waituntil statement.

In the rest of this paper, we assume the standard semantics of waituntil($p$){$s$} statements where a thread $t$ atomically performs the following actions: It first evaluates the boolean predicate $p$. If $p$ evaluates to *true*, $t$ also executes $s$ immediately after the evaluation of $p$. Otherwise, $t$ is blocked until $p$ becomes true.

Since the semantics of statements $s$ are standard, we do not present them in detail and use the notation $\langle s, t, \sigma \rangle \Downarrow \sigma'$ to indicate the resulting monitor state $\sigma'$ when thread $t$ executes statement $s$ under initial state $\sigma$. Given a monitor state $\sigma$, thread $t$, and predicate $p$, we write $(\sigma, t) \models p$ if $p$ evaluates to *true* and $(\sigma, t) \not\models p$ if $p$ evaluates to false.

**Monitor traces.** To define the semantics of monitors, we first introduce the notion of a *monitor trace*. A monitor trace $\tau$ is a sequence of *monitor events* where each event $e$ is a triple $(t, w, b)$ where $t$ is a thread identifier, $w$ is a waituntil statement, and $b$ is a boolean indicating whether the guard of $w$ evaluates to *true* or *false*. In particular, the event $(t, w, false)$ indicates that thread $t$ was blocked on the guard of $w$, whereas the event $(t, w, true)$ indicates that $t$ was able to execute $w$ in its entirety. Given an event $e = (t, w, b)$, we write $\overline{e}$ to denote the pair $(t, w)$.

We say that a monitor trace is *syntactically well-formed* if it (a) respects the relative ordering of statements within a method, (b) obeys the requirement that a thread cannot execute method $m'$ before finishing the execution of method $m$, and (c) satisfies the invariant that a thread exits the monitor either by blocking on a predicate or by finishing the

---

[2]i.e., condition c could have changed while the thread is waiting on p

$$
(1a) \quad \frac{\begin{array}{c} e = (t, w, \textit{false}) \\ \overline{e} \notin \mathcal{B} \quad (\sigma, t) \not\models \textit{Guard}(w) \end{array}}{(\sigma, e, \mathcal{B}, \mathcal{N}) \longrightarrow (\sigma, \epsilon, \mathcal{B} \cup \{\overline{e}\}, \mathcal{N})}
$$

$$
(1b) \quad \frac{\begin{array}{c} e = (t, w, \textit{false}) \\ \overline{e} \in \mathcal{N} \quad (\sigma, t) \not\models \textit{Guard}(w) \end{array}}{(\sigma, e, \mathcal{B}, \mathcal{N}) \longrightarrow (\sigma, \epsilon, \mathcal{B}, \mathcal{N} \setminus \{\overline{e}\})}
$$

$$
(2a) \quad \frac{\begin{array}{c} e = (t, w, \textit{true}) \\ \overline{e} \notin \mathcal{B} \quad (\sigma, t) \models \textit{Guard}(w) \\ \langle \textit{Body}(w), t, \sigma \rangle \Downarrow \sigma' \\ \mathcal{N}' = \{(t, w) \mid (t, w) \in \mathcal{B}, (\sigma', t) \models \textit{Guard}(w)\} \end{array}}{(\sigma, e, \mathcal{B}, \mathcal{N}) \longrightarrow (\sigma', \epsilon, \mathcal{B}, \mathcal{N} \cup \mathcal{N}')}
$$

$$
(2b) \quad \frac{\begin{array}{c} e = (t, w, \textit{true}) \\ \overline{e} = \min(\mathcal{N}) \quad (\sigma, t) \models \textit{Guard}(w) \\ \langle \textit{Body}(w), t, \sigma \rangle \Downarrow \sigma' \\ \mathcal{N}' = \{(t, w) \mid (t, w) \in \mathcal{B}, (\sigma', t) \models \textit{Guard}(w)\} \end{array}}{(\sigma, e, \mathcal{B}, \mathcal{N}) \longrightarrow (\sigma', \epsilon, \mathcal{B} \setminus \{\overline{e}\}, (\mathcal{N} \cup \mathcal{N}') \setminus \{\overline{e}\})}
$$

$$
(3) \quad \frac{(\sigma, e, \mathcal{B}, \mathcal{N}) \longrightarrow (\sigma', \epsilon, \mathcal{B}', \mathcal{N}')}{(\sigma, e :: \tau, \mathcal{B}, \mathcal{N}) \longrightarrow (\sigma', \tau, \mathcal{B}', \mathcal{N}')}
$$

**Figure 4.** Transition relation for implicit-signal monitor traces. Given an event $e = (t, w, b)$, $\overline{e}$ denotes $(t, w)$. We assume there is a total order relation $\prec$ between events and min picks the minimum one with respect to $\prec$.

$$
(1a) \quad \frac{\begin{array}{c} e = (t, w, \textit{false}) \\ \overline{e} \notin \mathcal{B} \quad (\sigma, t) \not\models \textit{Guard}(w) \end{array}}{(\sigma, e, \mathcal{B}, \mathcal{N}) \Longrightarrow (\sigma, \epsilon, \mathcal{B} \cup \{\overline{e}\}, \mathcal{N})}
$$

$$
(1b) \quad \frac{\begin{array}{c} e = (t, w, \textit{false}) \\ \overline{e} \in \mathcal{N} \quad (\sigma, t) \not\models \textit{Guard}(w) \end{array}}{(\sigma, e, \mathcal{B}, \mathcal{N}) \Longrightarrow (\sigma, \epsilon, \mathcal{B}, \mathcal{N} \setminus \{\overline{e}\})}
$$

$$
(2a) \quad \frac{\begin{array}{c} e = (t, w, \textit{true}) \\ \overline{e} \notin \mathcal{B} \quad (\sigma, t) \models \textit{Guard}(w) \\ \langle \textit{Body}(w), t, \sigma \rangle \Downarrow \sigma' \\ \mathcal{N}_1 = \textit{GetSignals}(w, \sigma', \mathcal{B}) \\ \mathcal{N}_2 = \textit{GetBroadcasts}(w, \sigma', \mathcal{B}) \end{array}}{(\sigma, e, \mathcal{B}, \mathcal{N}) \Longrightarrow (\sigma', \epsilon, \mathcal{B}, \mathcal{N} \cup \mathcal{N}_1 \cup \mathcal{N}_2)}
$$

$$
(2b) \quad \frac{\begin{array}{c} e = (t, w, \textit{true}) \\ \overline{e} = \min(\mathcal{N}) \quad (\sigma, t) \models \textit{Guard}(w) \\ \langle \textit{Body}(w), t, \sigma \rangle \Downarrow \sigma' \\ \mathcal{N}_1 = \textit{GetSignals}(w, \sigma', \mathcal{B}) \\ \mathcal{N}_2 = \textit{GetBroadcasts}(w, \sigma', \mathcal{B}) \end{array}}{(\sigma, e, \mathcal{B}, \mathcal{N}) \Longrightarrow (\sigma', \epsilon, \mathcal{B} \setminus \{\overline{e}\}, (\mathcal{N} \cup \mathcal{N}_1 \cup \mathcal{N}_2) \setminus \{\overline{e}\})}
$$

$$
(3) \quad \frac{(\sigma, e, \mathcal{B}, \mathcal{N}) \Longrightarrow (\sigma', \epsilon, \mathcal{B}', \mathcal{N}')}{(\sigma, e :: \tau, \mathcal{B}, \mathcal{N}) \Longrightarrow (\sigma', \tau, \mathcal{B}', \mathcal{N}')}
$$

**Figure 5.** Transition relation for explicit-signal monitor traces. Given an event $e = (t, w, b)$, $\overline{e}$ denotes $(t, w)$.

execution of a method. A more formal definition of syntactic well-formedness is presented in Appendix A.

**Example 3.2.** Consider the following monitor $M$, where we elide the "atomic" keywords for brevity:

```
monitor M {
  ...
  m1() {waituntil(x>0) {...}; waituntil{y>0}{...}}
  m2() {waituntil(z>0) {...}; waituntil{w>0}{...}}
}
```

Let us refer to the $j$'th waituntil statement in method $i$ as $w_{ij}$. The trace $[(1, w_{12}, \textit{true}), (1, w_{11}, \textit{true})]$ is not syntactically well-formed since the same thread cannot execute $w_{12}$ before $w_{11}$ (i.e., it violates requirement (a)). Similarly, the trace $[(1, w_{11}, \textit{false}), (1, w_{21}, \textit{true})]$ is also not syntactically well-formed since the same thread cannot execute method m2 before finishing the execution of m1 (violates (b)). Finally, the following trace is also not syntactically well-formed:

$$[(1, w_{11}, \textit{false}), (2, w_{21}, \textit{true}), (1, w_{11}, \textit{true}), (1, w_{12}, \textit{true})]$$

In particular, it violates requirement (c) since thread 2 exists the monitor without getting blocked or finishing the execution of m2. On the other hand, the following trace is syntactically well-formed:

$$[\ (1, w_{11}, \textit{false}), (2, w_{21}, \textit{true}), (2, w_{22}, \textit{false}),$$
$$(1, w_{11}, \textit{true}), (1, w_{12}, \textit{true}), (2, w_{22}, \textit{true})\ ]$$

In this trace, thread 1 attempts to execute the body of $w_{11}$ but is blocked (i.e., $x > 0$ evaluates to false). Then, thread 2 executes the first waituntil statement in method m2, but gets blocked on the second one. After thread 2 executes $w_{21}$, $x > 0$ becomes true, and thread 1 is able to finish executing method m1. Finally thread 2 finishes executing method m2.

**Semantics.** We now define the semantics of implicit-signal monitors in terms of the feasibility of well-formed monitor traces. Given a monitor $M$ and a monitor state $\sigma$, we say that a trace $\tau$ is *feasible* under $\sigma$ iff (a) it is syntactically well-formed and (b) $(\sigma, \tau, \emptyset, \emptyset) \longrightarrow^* (\sigma', \epsilon, \_, \_)$ where $\longrightarrow^*$ denotes the reflexive transitive closure of the transition relation $\longrightarrow$ defined in Figure 4.

Transition relations for implicit-signal monitors are described in Figure 4 using judgments of the form

$$(\sigma, \tau, \mathcal{B}, \mathcal{N}) \longrightarrow (\sigma', \tau', \mathcal{B}', \mathcal{N}')$$

where $\mathcal{B}$ and $\mathcal{N}$ describe blocked and notified threads respectively. In particular, $(t, w) \in \mathcal{B}$ indicates that thread $t$ is currently blocked on the predicate of $w$. In contrast, $(t, w) \in \mathcal{N}$ indicates that thread $t$ should be woken up to recheck the predicate of $w$. The meaning of the judgment $(\sigma, \tau, \mathcal{B}, \mathcal{N}) \longrightarrow (\sigma', \tau', \mathcal{B}', \mathcal{N}')$ is that executing the first event $e$ in $\tau$ under $\sigma$, $\mathcal{B}$, and $\mathcal{N}$ yields a new state $\sigma'$ as well as a new set of blocked and notified threads $\mathcal{B}'$ and $\mathcal{N}'$ respectively. We now explain the transition relations from Figure 4 in more detail.

According to rules (1a) and (1b), an event $e$ of the form $(t, w, \textit{false})$ is only feasible when $(\sigma, t) \not\models \textit{Guard}(w)$ (i.e., the predicate of $w$ evaluates to false). If $\overline{e} = (t, w)$ was not previously in the blocked thread set $\mathcal{B}$, rule (1a) adds $\overline{e}$ to $\mathcal{B}$.

$$
\begin{aligned}
\textit{Events}(\mathcal{B}, p) &= \{(t, w) \mid (t, w) \in \mathcal{B} \land \textit{Guard}(w) = p\} \\
\textit{GetSignals}(w, \sigma, \mathcal{B}) &= \{(t', w') \mid (p, c) \in \textit{Signals}(w) \land (t', w') = \min(\textit{Events}(\mathcal{B}, p)) \land (c = \checkmark \lor (\sigma, t') \models p)\} \\
\textit{GetBroadcasts}(w, \sigma, \mathcal{B}) &= \{(t', w') \mid (p, c) \in \textit{Broadcasts}(w) \land (t', w') \in \textit{Events}(\mathcal{B}, p) \land (c = \checkmark \lor (\sigma, t') \models p)\}
\end{aligned}
$$

**Figure 6.** Auxiliary functions used in Figure 5

If $\overline{e}$ was already in $\mathcal{B}$, then $e$ is only feasible if $\overline{e}$ was "notified" by the system (i.e., $\overline{e} \in \mathcal{N}$ in rule (1b)).

The next two rules (2a) and (2b) state that an event $e = (t, w, \textit{true})$ is only feasible when $(\sigma, t) \models \textit{Guard}(w)$ (i.e., the predicate of $w$ evaluates to true). Both rules execute the body of $w$ to obtain a new monitor state $\sigma'$. Now, since the execution of $w$ may cause the predicates of blocked threads to become true, $\mathcal{N}'$ contains all $(t, w)$ pairs that were previously in $\mathcal{B}$ and whose predicates evaluate to true under $\sigma'$.

### 3.3 Target Language

Our target language is very similar to the source language from Figure 3, except that the body of waituntil statements contain explicit signals. In particular, a waituntil construct in the target language looks as follows:

```
waituntil(p){ s; signal(S₁); broadcast(S₂) }
```

Here, $S_1$ and $S_2$ are sets of pairs $(p, c)$ where $p$ is a predicate and $c \in \{?, \checkmark\}$. The informal semantics of signal and broadcast are as follows: If $(p, \checkmark)$ is in $S_1$, then the system will notify (i.e., wake up) a *single* thread blocked on predicate $p$. In contrast, if $(p, \checkmark) \in S_2$, then the system notifies *all* threads blocked on $p$. On the other hand, if $(p, ?)$ is in $S_1$ (resp. $S_2$), then $p$ will be evaluated at run-time, and, if $p$ evaluates to true, then one thread (resp. all threads) blocked on $p$ will be notified. Given a waituntil statement in the target language, we write $\textit{Signals}(w)$ to indicate $S_1$ and $\textit{Broadcasts}(w)$ to represent $S_2$.

We also describe the formal semantics of explicit-signal monitors in terms of monitor traces, where the definitions of *trace*, *event*, and *well-formedness* remain the same as in Section 3.2. However, the concept of *feasibility* is defined with respect to a different transition relation $\Longrightarrow$, shown in Figure 5. In particular, we say that an explicit-signal monitor trace $\tau$ is *feasible* if (a) it is syntactically well-formed, and (b) $(\sigma, \tau, \emptyset, \emptyset) \Longrightarrow^* (\sigma', \epsilon, \_, \_)$ where $\Longrightarrow^*$ denotes the reflexive transitive closure of the transition relation $\Longrightarrow$ from Figure 5.

The transition relation $\Longrightarrow$ is defined similarly as $\longrightarrow$ except for events of the form $(t, w, \textit{true})$. In contrast to implicit signal monitors which wake up all threads whose predicates have become true, explicit-signal monitors decide which threads to notify based on $\textit{Signals}(w)$ and $\textit{Broadcasts}(w)$. In particular, rules (2a) and (2b) use auxiliary functions $\textit{GetSignals}$ and $\textit{GetBroadcasts}$ (defined in Figure 6) to decide which threads to add to the notification set $\mathcal{N}$. If $(p, c) \in \textit{Signals}(w)$, then we notify a *single* event $(t', w') \in \mathcal{B}$ such that the predicate of $w'$ is $p$. If $c = ?$, we additionally check that $p$ evaluates to true under $\sigma'$ before adding $(t', w')$ to the notification set. The function $\textit{GetBroadcasts}$ is defined similarly except that it notifies all threads blocked on the specified predicate rather than a single one.

### 3.4 Equivalence

We are now ready to define what it means for an implicit-signal monitor $M$ from the source language and an explicit-signal monitor $M'$ from the target language to be *equivalent*. Towards this goal, we first define a normal form for traces:

**Definition 3.3. (Normalization)** Let $\tau$ be an implicit-signal monitor trace. We say that $\tau$ is *normalized* with respect to monitor state $\sigma$ if we can derive $(\sigma, \tau, \emptyset, \emptyset) \longrightarrow^* (\sigma, \epsilon, \_, \_)$ without using rule (1b) from Figure 4 in the derivation.

Since rule (1b) corresponds to a spurious notification, a trace is normalized if threads are woken up only when their predicates evaluate to true. Observe that we can always find a normalized feasible trace for any feasible trace by changing the order in which threads are woken up.[3]

**Definition 3.4. (Equivalence)** Let $M, M'$ be implicit- and explicit-signal monitors respectively. We say that $M$ and $M'$ are semantically equivalent, written $M \sim M'$, iff for all monitor states $\sigma$ and all well-formed traces $\tau$, the following two conditions are satisfied:

1. If $(\sigma, \tau, \emptyset, \emptyset) \Longrightarrow^* (\sigma', \epsilon, \_, \_)$, then it is also the case that $(\sigma, \tau, \emptyset, \emptyset) \longrightarrow^* (\sigma', \epsilon, \_, \_)$.
2. If $(\sigma, \tau, \emptyset, \emptyset) \longrightarrow^* (\sigma', \epsilon, \_, \_)$ and $\tau$ is normalized with respect to $\sigma$, then $(\sigma, \tau, \emptyset, \emptyset) \Longrightarrow^* (\sigma', \epsilon, \_, \_)$.

Here, the first condition states that any feasible trace of the explicit-signal monitor $M'$ must also be a feasible trace of its implicit version $M$. However, in general, we cannot expect the converse of this statement to hold: Since the explicit-signal monitor may be more efficient than its implicit-signal counterpart, we cannot require that all feasible traces of $M$ to be also feasible in the explicit-monitor case. Thus, the second condition states that any *normalized* feasible trace of $M$ should also be feasible in $M'$.

## 4 Signal Placement Algorithm

In this section, we describe our algorithm for automatically transforming an implicit-signal monitor $M$ in the source language to an explicit-signal monitor $M'$ in the target language. Our algorithm ensures that $M$ and $M'$ are equivalent in the sense of Definition 3.4 and also tries to minimize the number of spurious wake-ups and conditional signals in $M'$. We

---

[3]Recall that our notion of feasibility does not require the sets $\mathcal{B}, \mathcal{N}$ to be empty. In particular, a trace $\tau$ is feasible under $\sigma$ if $(\sigma, \tau, \emptyset, \emptyset) \longrightarrow^* (\sigma', \epsilon, \mathcal{B}, \mathcal{N})$ for any $\mathcal{B}$ and $\mathcal{N}$. Therefore, a notification that would have been eliminated by rule (1b) can just be ignored indefinitely, i.e., remain in the $\mathcal{N}$ set without affecting other transitions.

**Algorithm 1** Signal Placement Algorithm

1: **function** PLACESIGNALS(M, I)
2:    **input:** $M$, an implicit signal monitor
3:    **input:** $I$, a monitor invariant
4:    **output:** $M'$, an explicit signal monitor
5:    $\Sigma \leftarrow [w \mapsto \emptyset \mid w \in CCRs(M)]$
6:    **for** $(w, p) \in CCRs(M) \times Guards(M)$ **do**
7:      **if** $\vdash \{I \wedge Guard(w) \wedge \neg p\}\ Body(w)\ \{\neg p\}$ :
8:        **continue**;
9:      **if** $\vdash \{I \wedge Guard(w) \wedge \neg p\}\ Body(w)\{p\}$ :
10:        $cond \leftarrow \checkmark$
11:      **else**
12:        $cond \leftarrow ?$
13:      **if** $\forall (p, s') \in CCRs(M).\ \vdash \{I \wedge p\}\ s'\{\neg p\}$ :
14:        $bcast \leftarrow false$
15:      **else**
16:        $bcast \leftarrow true$
17:      $\Sigma(w) \leftarrow \Sigma(w) \cup \{(p, cond, bcast)\}$
18:    **return** Instrument($M, \Sigma$)

start with a basic version of the algorithm and then describe extensions and improvements later in this section.

### 4.1 Basic Algorithm

Our basic signal placement algorithm is shown in Algorithm 1. The PLACESIGNALS algorithm takes as input an implicit-signal monitor $M$ as well as a *monitor invariant* $I$, which is an assertion that holds every time a thread enters and exits the monitor. Since automated inference of monitor invariants is described in the next section, we will assume that an oracle provides them for the time being. In this section, we further assume that guards used in waituntil statements do not contain thread-local variables. Given such an implicit-signal monitor $M$ and its invariant $I$, PLACESIGNALS returns an explicit-signal monitor $M'$ such that $M \sim M'$.

The algorithm maintains a mapping from each *conditional critical region (CCR)* (i.e., waituntil statement) $w$ in $M$ to a set of notifications that should be performed after executing $w$ and before exiting the monitor. The algorithm represents these notification as triples of the form $(p, cond, bcast)$, where $p$ is a predicate, $cond \in \{?, \checkmark\}$ indicates whether the notification is conditional or unconditional, and $bcast$ is a boolean indicating whether it is necessary to notify all threads blocked on $p$ as opposed to a single one. Once the algorithm computes this mapping $\Sigma$, it instruments the original implicit-signal monitor $M$ as shown in Figure 7 to obtain an explicit-signal monitor $M'$.

The key part of the PLACESIGNALS algorithm is the loop in lines 6–17. For each conditional critical region $w$ and predicate $p$ used in the monitor, the algorithm first decides whether $w$ may need to notify threads blocked on predicate $p$. This decision is made based on the provability of the

$$\frac{\begin{array}{c}w = \texttt{waituntil}(p')\{s\}\\ S_1 = \{(p, c) \mid (p, c, false) \in \Sigma(w)\}\\ S_2 = \{(p, c) \mid (p, c, true) \in \Sigma(w)\}\\ s' = \texttt{signal}(S_1); \texttt{broadcast}(S_2)\end{array}}{\Sigma \vdash w \rightsquigarrow \texttt{waituntil}(p')\{s; s'\}}$$

$$\frac{\Sigma \vdash w_1 \rightsquigarrow w'_1 \quad \ldots \quad \Sigma \vdash w_n \rightsquigarrow w'_n}{\Sigma \vdash w_1; \ldots; w_n \rightsquigarrow w'_1; \ldots; w'_n}$$

**Figure 7.** Performing instrumentation

following Hoare triple:

$$\{I \wedge Guard(w) \wedge \neg p\}\ Body(w)\ \{\neg p\}$$

Essentially, this triple says that executing the body of $w$ in a state in which $\neg p$ holds ensures that predicate $p$ continues to remain false. Hence, any thread $t$ blocked on $p$ will remain blocked after executing $w$, so there is no need to notify $t$. Observe that the precondition of the Hoare triples also assumes $I \wedge Guard(w)$ because (a) $Guard(w)$ is a prerequisite for executing the body of $w$ and, (b) by definition of monitor invariant, $I$ must hold before executing the body of any CCR.

Next, lines 9–12 determine whether the notification should be conditional or not. Recall that a conditional notification for predicate $p$ checks whether $p$ evaluates to true before waking up threads blocked on $p$. While conditional notifications prevent spurious wake-ups, it is desirable to avoid evaluating $p$ at run-time if $p$ is guaranteed to hold after executing $w$. Thus, line 9 checks the validity of the following Hoare triple:

$$\{I \wedge Guard(w) \wedge \neg p\}\ Body(w)\ \{p\}$$

In other words, assuming we execute $w$ in a state where a thread is blocked on $p$, the execution of $Body(w)$ results in a state where $p$ is true. Thus, there is no need to evaluate $p$ at run time before signaling threads blocked on $p$.

The last part of Algorithm 1 (lines 13–16) determines whether we should notify *all* threads blocked on predicate $p$. Suppose there are $n$ threads $T = \{t_1, \ldots, t_n\}$ blocked on $p$, and suppose that an arbitrary thread $t_i$ gets unblocked. If executing $t_i$ is guaranteed to result in a state where predicate $p$ is false, then it is not necessary to notify any of the remaining threads $T \setminus \{t_i\}$. Thus, the algorithm checks the following Hoare triple for *all* CCRs $w'$ with guard $p$:

$$\{I \wedge p\}\ Body(w')\ \{\neg p\}$$

If this Hoare triple holds for all CCRs with guard $p$, then it is safe to signal rather than broadcast.

**Theorem 4.1.** [4] *Let* $PlaceSignals(M, I) = M'$. *If $I$ is a correct monitor invariant and guards of CCRs in $M$ do not contain thread-local variables, then $M \sim M'$.*

---
[4]The proofs of all theorems are in the appendix.

## 4.2 Handling Thread-Local Variables

To simplify presentation, our algorithm from Section 4.1 assumes that guards of CCRs in the input monitor do not contain thread-local variables. However, if the input monitor $M$ does not satisfy this assumption, the explicit-signal monitor $M'$ generated by Algorithm 1 may not be equivalent to $M$. We illustrate the problem using the following example:

**Example 4.2.** Consider the following monitor:
```
monitor M {
  int y=0;
  m1(int x) {waituntil(x < y) {x = y+1;} }
  m2() { y = y+2; } }
```

Suppose we have threads $t_1, t_2, t_3$, where $t_1, t_2$ are blocked in m1, and $t_3$ is executing m2, after which the value of y becomes 2. Further, suppose that the value of the thread-local variable x is 0 for $t_1$ and 1 for $t_2$. Since the predicate $x < y$ has become true for both $t_1, t_2$ and executing $t_1$ does not change the value of the predicate in $t_2$ (and vice versa), $t_3$ should notify both threads. Thus, the explicit-signal monitor should use broadcast instead of signal.

However, recall that Algorithm 1 determines whether m2 should broadcast or signal by checking the validity of $\{x < y\}$ $x = y + 1$ $\{x \geq y\}$. Since this Hoare triple is valid, we would erroneously conclude that it is safe for m2 to notify a single thread instead of all threads.

As illustrated by this example, Algorithm 1 is unsound when guards contain thread-local variables. To remedy this situation, we need to rename thread-local variables when checking validity. In particular, recall that PlaceSignals checks the validity of Hoare triples of the form $\{P_1 \wedge P_2\} S \{Q\}$ where $P_1$ is an assumption about the currently running thread, whereas $P_2$ and $Q$ are assumptions/assertions about some other thread. Since $S$ and $P_1$ may refer to thread-local variables that are also used in $P_2$ and $Q$, we need to rename thread-local variables and check the validity of the following modified Hoare triple:

$$\{P_1 \wedge P_2[V'/V]\} S \{Q[V'/V]\}$$

where $V = Locals(P_2) \cup Locals(Q)$ and $V'$ denotes a fresh set of variables not used elsewhere in $P_1, P_2, S$, and $Q$.[5]

## 4.3 Improvement over the Basic Algorithm

In this section, we consider an improvement over Algorithm 1 that aims to further reduce the number of broadcasts in the synthesized explicit-signal monitor. Recall that Algorithm 1 determines whether a CCR should notify one vs. all threads blocked on predicate $p$ by checking the validity of the following Hoare triple for all CCRs $w$ with guard $p$:

$$\{I \wedge p\} \; Body(w) \; \{\neg p\} \tag{1}$$

In some cases, it is possible to further strengthen the precondition of this Hoare triple. In particular, suppose that the signaling CCR is $w'$ with body $s'$ and guard $p'$ and suppose that $\phi$ is guaranteed to hold after executing $s'$. In general, we cannot assume $\phi$ in the pre-condition of Equation 1 because other threads may have invalidated $\phi$ before the notified thread has a chance to execute. However, if $s$ commutes with the body of every other CCR in the monitor, then the monitor state after executing $s$ for *any* interleaving is equivalent to one in which we execute $s$ immediately after $s'$. In this case, we can safely assume that $s$ executes immediately after $s'$ since the resulting states are equivalent. This insight allows us to strengthen the precondition of Equation 1 by using the post-condition $\phi$ of the signaling thread.

---
[5]Another subtlety with local variables is how to signal conditionally. We discuss evaluation of predicates with local variables in Section 6.

To make this discussion more precise, let us define a predicate $Comm(w, M)$ as follows:

$$Comm(w, M) \Leftrightarrow \big(\forall w' \in CCRs(M) \setminus \{w\}.\\ Body(w'); Body(w) \equiv Body(w); Body(w')\big)$$

Essentially, this predicate is true if the body of $w$ commutes with every other CCR in the monitor. Now, using this definition, we can state a weaker sufficient condition for CCR $w$ to notify one –rather than all– threads blocked on predicate $p$. In particular, we can change the condition at line 13 of Algorithm 1 to the following weaker one:

$$\forall w' = (p, s') \in CCRs(M). \big( \vdash \{I \wedge p\}s'\{\neg p\} \vee \\ (Comm(w', M) \wedge \vdash \{I \wedge Guard(w) \wedge \neg p\}Body(w); s'\{\neg p\})\big) \tag{2}$$

The first line of Equation 2 corresponds to the same check we perform at line 13 in Algorithm 1 to determine whether it is safe to signal rather than broadcast. However, if this condition does not hold, we may still be able to prove that broadcasting is unnecessary as long as $s'$ commutes with every other CCR in the monitor and we can prove that $p$ is falsified after executing $Body(w); s'$.

The correctness of Equation 2 follows from the following theorem (and the proof of Theorem 4.1):

**Theorem 4.3.** *Let $\tau = \tau_0 e$ be a monitor trace and let $\tau' = e\tau_0$ where $e = (t, w, b)$. If $(\sigma, \tau, \mathcal{B}, \mathcal{N}) \longrightarrow^* (\sigma', \epsilon, \mathcal{B}', \mathcal{N}')$ and $Comm(w, M)$, then we have $(\sigma, \tau', \mathcal{B}, \mathcal{N}) \longrightarrow^* (\sigma', \epsilon, \mathcal{B}', \mathcal{N}')$.*

**Remark.** Our discussion in this section assumes non-preemptive signal semantics [3] where a signaled thread is not guaranteed to consume the signal immediately. However, if we assume *preemptive* signal semantics, we can perform this optimization more liberally by only checking whether predicate $p$ is invalidated by the sequential composition of the segment that produces the signal and $Body(w)$.

## 5 Inference of Monitor Invariants

Our signal placement algorithm from Section 4 relies on a monitor invariant $I$ that holds at the entry and exit of every CCR. In this section, we describe our method for automatically inferring useful monitor invariants.

Our inference algorithm is property-directed in that it only infers invariants that are useful for proving the Hoare triples generated by the signal placement algorithm. Specifically,

**Algorithm 2** Monitor Invariant Inference

1: **function** INFERMONITORINV($M, \Theta$)
2:    **input:** $M$, an implicit signal monitor
3:    **input:** $\Theta$, set of Hoare triples of the form $\{P\}\ s\ \{Q\}$
4:    **output:** $I$, a monitor invariant
5:    $\Phi \leftarrow \emptyset$
6:    **for** $\{P\}\ s\ \{Q\} \in \Theta$ **do**
7:       $\Phi \leftarrow \Phi \cup abduce(P, \mathsf{wp}(s, Q))$
8:    **do**
9:       $numPreds \leftarrow |\Phi|$
10:      **for** $\psi \in \Phi$ **do**
11:         **if** $\nvdash \{true\}\ Ctr(M)\ \{\psi\}$ :
12:           $\Phi \leftarrow \Phi \setminus \{\psi\}$
13:           **continue**;
14:         $I \leftarrow \bigwedge_{\psi_i \in \Phi} \psi_i$
15:         **if** $\exists w \in CCRs(M). \nvdash \{I \land Guard(w)\}\ Body(w)\ \{\psi\}$ :
16:           $\Phi \leftarrow \Phi \setminus \{\psi\}$
17:    **while** $numPreds \neq |\Phi|$
18:    **return** $I$

our inference engine uses *abductive reasoning* [16] to automatically infer predicates that are useful for proving a given set of Hoare triples. Given a universe of predicates $\Phi$ generated using abduction, it then infers the *strongest* conjunctive monitor invariant over predicates in $\Phi$. Therefore, our invariant inference engine can be viewed as marrying the power of abductive reasoning with predicate abstraction [21, 37]. The advantage of this approach is two-fold: First, rather than relying on a hard-coded universe of predicate templates, our algorithm infers useful predicates automatically using abduction. Second, because the predicates inferred using abduction can involve disjunctions, the monitor invariants synthesized by our algorithm are not restricted to pure conjunctions.

With this intuition in mind, we now explain our INFERMONITORINV procedure from Algorithm 2 in more detail. This procedure takes two inputs, namely, an implicit-signal monitor $M$ and a set $\Theta$ of Hoare triples of the form $\{P\}\ s\ \{Q\}$. Note that $\Theta$ simply corresponds to the set of Hoare triples generated by Algorithm 1, but with $I$ set to *true*. The return value of INFERMONITORINV is a formula $I$ representing a valid monitor invariant of $M$.

Conceptually, the INFERMONITORINV procedure operates in two phases. The first phase (lines 5–7) generates a universe $\Phi$ of candidate predicates, and the second phase (lines 8–17) performs fixed-point computation to infer the strongest conjunctive monitor invariant $I$ over predicates in $\Phi$.

In the first phase of the algorithm, we iterate over all Hoare triples $\{P\}\ s\ \{Q\}$ in $\Theta$ and look for a strengthening $\psi$ of the precondition such that the Hoare triple $\{P \land \psi\}\ s\ \{Q\}$ becomes valid. Because the correctness of the Hoare triple $\{P\}\ s\ \{Q\}$ boils down to checking the validity of the formula $P \Rightarrow \mathsf{wp}(s, Q)$, we can find a suitable strengthening of $P$ by solving the following abductive reasoning problem:

$$\text{Find } \psi \text{ such that :} \quad (1)\ P \land \psi \models \mathsf{wp}(s, Q) \quad (2)\ P \land \psi \nvDash \mathit{false} \quad (3)$$

Here condition (1) states that $\{P \land \psi\}\ s\ \{Q\}$ is a valid Hoare triple, and (2) states that the speculated invariant $\psi$ is consistent with precondition $P$. Since abductive reasoning is a well-studied problem, we use the *abduce* procedure described in prior work [16] to automatically infer candidate strengthenings $\psi$. Also, note that a call to *abduce* at line 7 may yield multiple predicates $\psi_1, \ldots, \psi_n$, all of which constitute valid solutions for Equation 3.

Since the predicates $\Phi$ generated using abduction in lines 5–7 are merely *candidate* invariants, the next phase of the algorithm performs a fixed-point computation in which we drop every $\psi \in \Phi$ that is not a monitor invariant. Specifically, for each predicate $\psi$ in $\Phi$, we check whether (a) it holds initially (lines 11–13) and (b) whether it is preserved by each CCR in the monitor (lines 15–16). To determine whether $\psi$ holds initially, we check the validity of the Hoare triple $\{true\}\ Ctr(M)\ \{\psi\}$, where $Ctr(M)$ denotes the constructor of $M$.[6] If this Hoare triple is not valid, we simply drop $\psi$ from set $\Phi$. Next, to determine whether $\psi$ is preserved by CCR $w$, we check the validity of the Hoare triple $\{I \land Guard(w)\}\ Body(w)\ \{\psi\}$, where $I$ denotes the conjunction of all predicates in $\Phi$. If this triple is invalid for any CCR in $M$, we again drop $\psi$ from the set $\Phi$. We then repeat this process until $I$ satisfies both the initiation and consecution requirements. It is easy to see that formula $I$ returned by INFERMONITORINV constitutes a valid monitor invariant.

## 6 Implementation

We have implemented our proposed method in a tool called EXPRESSO. Our implementation leverages the Soot program analysis infrastructure [46] and invokes the Z3 SMT solver [12] for checking logical validity. In what follows, we discuss some important design choices that are not addressed in previous sections.

***Generating Java code.*** While the target language (IR) presented in Section 3.3 is convenient for describing our transformation, it does not yield valid Java code. Our implementation converts programs in this IR to valid Java code in the following manner. First, we associate a condition variable with the guard of every waituntil statement. Now, given a waituntil statement $w$ with associated guard $p$, body $s$, and condition variable $c$, we then generate the following code[7]:

    **while**(!p) {c.await();};  s

Furthermore, for each $(p_i, ?) \in \mathit{Signals}(w)$, we generate the code if($p_i$) $c_i$.signal(), and for $(p_i, \checkmark) \in \mathit{Signals}(w)$, we emit $c_i$.signal(). For each $(p_i, \_) \in \mathit{Broadcasts}(w)$, we generate the same code where signal is replaced with signalAll.

---

[6]For simplicity, we assume a single constructor; if there are multiple ones, this triple needs to be checked for all constructors.
[7]Note that our implementation uses the ReentrantLock class.

***Instrumentation for predicates with local variables.*** To support conditional signaling for predicates with local variables, Expresso augments the monitor code with a data structure that tracks the values of local variables for any thread that is blocked on a predicate $p$. The code generated by Expresso then uses this data structure to check whether $p$ actually evaluates to true at program points that require conditional signaling for $p$.

***Lazy broadcasts.*** Expresso provides an option for performing broadcasts lazily. Consider a `waituntil` statement $w$ such that $(p, \_) \in \textit{Broadcasts}(w)$. Rather than emitting the code `c.signalAll()` after the body of $w$, "lazy broadcast" notifies a single thread $t$ blocked on $p$ and ensures that $t$ notifies all other threads by adding the instrumentation `if(p) c.signal()` after every waituntil statement with guard $p$. In our implementation, we enable this option by default to minimize context switches.

***Discharging Hoare triples.*** Expresso discharges any Hoare triple $\{P\}\ s\ \{Q\}$ by computing the weakest precondition of $Q$ with respect to $s$ and perfoming a validity check. Since $s$ can contain pointers, Expresso uses the points-to information provided by Doop [6] to produce a whole-program model of the heap. In particular, given a store statement $v.f = e$, Expresso generates additional statements of the form `if`$(v = x_i)\ x_i.f = e$ where $x_i$ is a potential alias of $v$.

## 7 Evaluation

We evaluate Expresso by performing experiments that are designed to answer the following research questions:

- How does the code generated by Expresso compare against hand-written explicit-signal code?
- How does our solution compare against run-time systems that provide support for implicit signals?
- How long does Expresso take to generate code?

***Benchmarks.*** The benchmarks used in our evaluation come from two different sources, namely all AutoSynch benchmarks from [30] and monitors collected from popular opensource projects from Github. We collected the Github benchmarks by writing a crawler that identifies potential monitors in Java programs using keywords such as `wait`, `signal`, `notify` etc. We then manually inspected these results in decreasing order of Github ranking (a mix of stars and forks) and identified self-contained modules (i.e., monitors) that encapsulate shared state. This process requires manual effort because we need to isolate the monitor code and insert it in a stress-testing harness.

***Performance evaluation methodology.*** We evaluate performance using the same methodology used for evaluating AutoSynch in [30]. Specifically, we use *saturation tests* [8] in which threads only access the monitor and perform no extra work outside of the monitor. This set-up allows us to stress-test the monitor code and meaningfully compare our solution with run-time solutions and near-optimal handwritten code (from the original AutoSynch benchmarks or from the GitHub project).

We perform measurements using the JMH framework [44], which is a benchmarking tool for rigorous measurement in JVM-based languages. All measurements are conducted on a 16-way (8 core x 2 SMT) Intel Xeon CPU E5-2640 v3 2.60GHz with 132 GB of memory using JDK 1.8.0_101-b13.

***Performance results.*** The results of our performance evaluation are presented in Figures 8 and 9. Specifically, Figure 8 shows the results for the AutoSynch benchmarks, augmented with the readers-writers example of Section 2. Figure 9 presents results for monitors found in popular GitHub projects. Each graph plots the average time (in milliseconds) per monitor operation (e.g., `enterReader`) against the number of threads.

In virtually all cases, the performance of Expressogenerated code is very close to hand-written explicit-signal code. The only significant differences are in the "H2O Barrier" benchmark under low concurrency and "Dining Philosophers" under high concurrency. In the latter, the explicit signalling code has knowledge of the problem structure itself, so it avoids all wakeups that do not lead to progress.

Comparing to AutoSynch, Expresso outperforms it by 1.56x on average over all benchmarks. Expresso significantly outperforms AutoSynch for about half the benchmarks of Figure 8, which are chosen or written by the AutoSynch authors themselves. On a few occasions, AutoSynch slightly outperforms Expresso-generated code. As we discuss in Section 8, AutoSynch offers dynamic structures for quick inequality comparisons between shared variables and local values (which are captured as constants while the thread is waiting). This custom optimization can also be added to Expresso but the emphasis of our work has been on statically eliminating unnecessary signalling, rather than minimizing the overhead of dynamic checks.

The monitors found in GitHub projects (Figure 9) are more representative of synchronization patterns in-the-wild. Expresso performs very well on these benchmarks, matching hand-optimized code and significantly outperforming AutoSynch: by 1.62x on average, and up to 2.5x on a high-concurrency setting with 128 threads.

Upon closer inspection of these benchmarks, we observe that the symbolic reasoning needed to achieve the results from Figure 9 is far from trivial. As a simple example, "ConcurrencyThrottle" from the Spring framework has a waiting condition `threadCount < threadLimit` triggered by the statement `threadCount--` in the monitor exit operation. In order to avoid broadcasts, Expresso needs to infer a monitor invariant that allows it to establish that whenever a thread enters the monitor, the waiting condition has to become true again due to a `threadCount++` operation. Because these increment and decrement operations are distant, symbolic reasoning has to model the semantics of all intervening program statements and establish that the operations commute. This kind of reasoning is necessary for Expresso to achieve the performance results from Figure 9 in all benchmarks. Furthermore, the inferred monitor invariants are often intricate—for instance, the "AsyncDispatch" invariant (shown in Appendix D) from the Gradle codebase has 22 sub-terms (12 equality/inequality comparisons and arithmetic operations and 10 logical connectives).

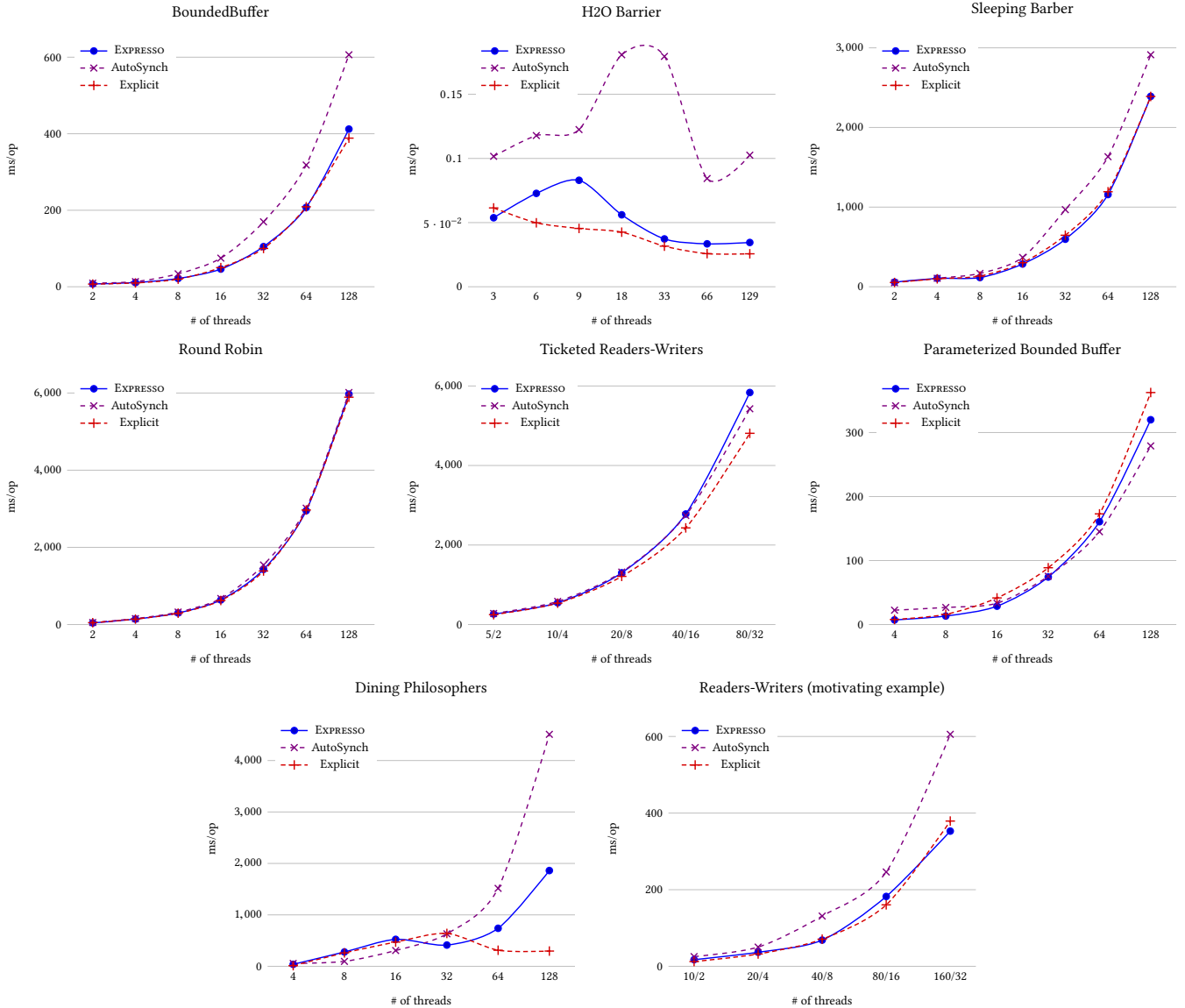

**Figure 8.** Performance over AutoSynch benchmarks and readers-writers example.

To summarize, these results demonstrate the plausibility of a practical and efficient implementation for implicit-signal monitors. In particular, the code generated by Expresso is comparable to hand-written code even for saturation tests that stress-test the monitor. Furthermore, Expresso's implicit-signal monitor implementation consistently outperforms AutoSynch on monitors extacted from real-world codebases such as Spring framework, Gradle, ExoPlayer, greenDAO, etc.

***Analysis time.*** Table 1 shows the time that Expresso takes to synthesize the explicit-signal code from its corresponding implicit-signal version for each benchmark. In most cases, the symbolic reasoning time is in the order of a few seconds. The only exception is the largest benchmark, AsyncDispatch, whose compilation takes 28.3 seconds. This example takes longer to analyze because some of the predicates depend on Java library operations that Expresso also needs to analyze. Overall, these results demonstrate that Expresso is practical and that it generates code whose performance is comparable to hand-written code in virtually all cases.

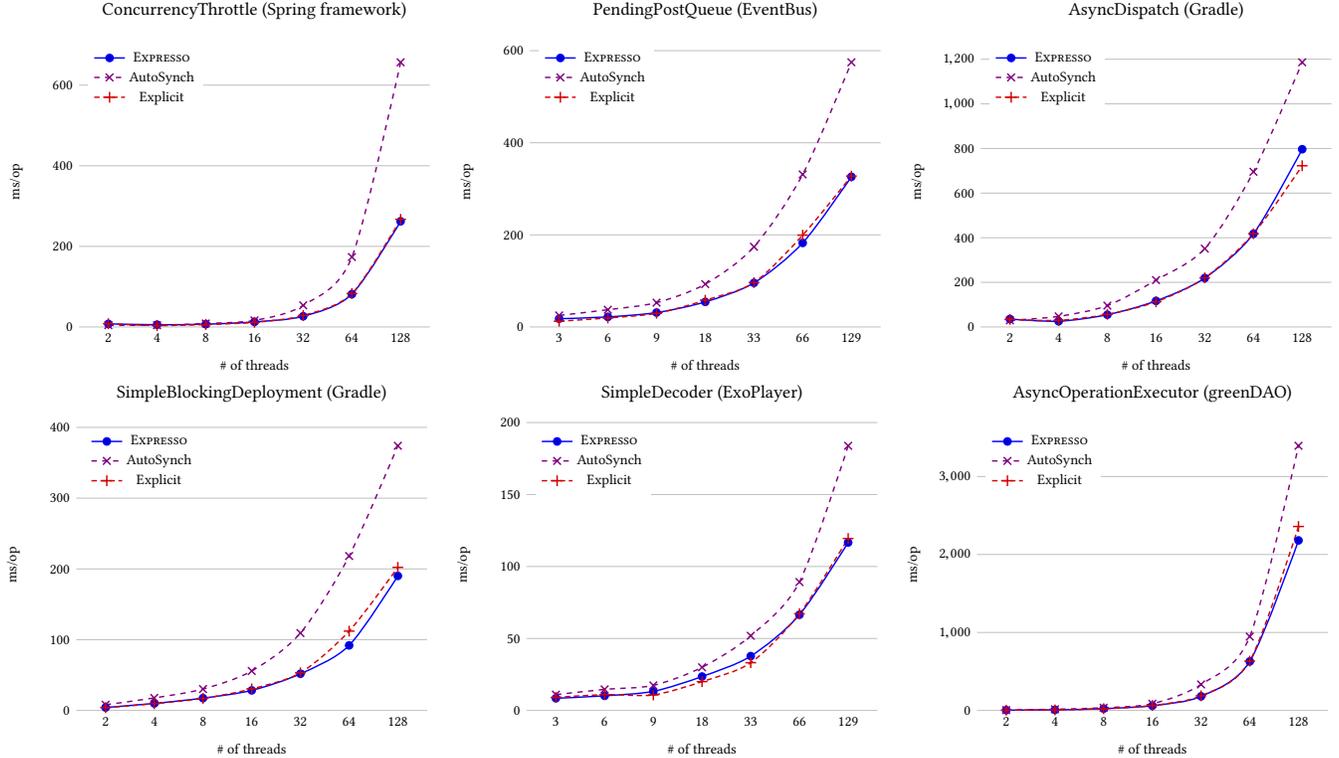

**Figure 9.** Performance over monitor code in popular GitHub projects.

***Generated code.*** We also assessed the quality of the code generated by Expresso by manually inspecting the synthesized explicit-signal monitors. For most benchmarks obtained from Github projects, we found that the code generated by Expresso is very similar to hand-written code. In the case of the AutoSynch benchmarks, however, we found some examples (e.g., Dining Philosophers) where the Expresso-generated code differs significantly from hand-written code. For these benchmarks, manually-written code leverages dynamic data structures to achieve optimal signaling, whereas Expresso uses the fixed strategy described in Section 6 for handling predicates with local variables.

## 8  Related Work

Our implicit-signal monitors have several relatives in the literature. We discuss representative past work, in loose thematic groupings of decreasing affinity with our work.

***Language designs and run-time support for concurrency.*** There is a very rich literature on language support for concurrency, dating back to the early 1960s [3]. Dijkstra originally proposed the concept of *semaphores* to provide a friendlier and more efficient programming abstraction than the busy-wait design [15]. In later work, Hoare proposed the concept of *conditional critical regions (CCR)* [27], which overcome some of the difficulties associated with semaphores by providing a more structured notation for specifying synchronization. In particular, every shared variable in a CCR must belong to a resource, and variables in a resource can only be accessed within so-called *region* statements of the form region r when B → S, where *B* is a guard and *S* is a statement. Concurrently to the introduction of CCRs, Dijkstra proposed the notion of *monitors*, which provide more structure than conditional critical regions and can be implemented as efficiently as semaphores [14]. To this day, monitors remain a popular concurrent programming paradigm, and the "monitor pattern" is widely used in many programming languages, including Java and C++.

| Benchmark | Time (sec.) |
|---|---|
| BoundedBuffer | 2.5 |
| H2OBarrier | 2.3 |
| Sleeping Barber | 1.6 |
| Round Robin | 1.2 |
| Ticketed Readers-Writers | 3.8 |
| Param. Bounded Buffer | 2.5 |
| Dining Philosophers | 5.4 |
| Readers-Writers | 1.5 |
| ConcurrencyThrottle | 1.0 |
| PendingPostQueue | 0.5 |
| AsyncDispatch | 28.3 |
| SimpleBlockingDeployment | 0.4 |
| SimpleDecoder | 10.7 |
| AsyncOperationExecutor | 2.1 |

**Table 1.** Compilation time for benchmarks.

Even though early proposals for monitors advocate an implicit signaling mechanism [3, 4], most modern monitor implementations use explicit signaling due to performance considerations. More recent work in this area aim to popularize implicit signal monitors by providing a more efficient implementation [8, 30]. For example, the recent AutoSynch work [30] attempts to improve the efficiency of automatic signaling through a combination of efficient dynamic indexing and simple static analysis. AutoSynch offers sophisticated handling of local state in a thread. If threads wait on conditions based on standard equality/inequality patterns over local variables, the system dynamically snapshots the values of local variables and treats them as run-time constants—the variable values cannot have changed while the thread is waiting. As a result, the predicates have a standard structure of comparisons with constants. The efficient notification algorithm then works much like database indexing: it computes, given which shared values changed, what waiting predicates could possibly have been affected. Our approach is distinguished by its use of reasoning techniques for statically inferring when a condition must, may-not, or may have become true. In order to do so, our technique needs to take into consideration several issues that are not addressed by previous work, e.g., handling of memory aliases, inter-procedural reasoning, etc. As shown in the evaluation, our method significantly outperforms the AutoSynch solution on many benchmarks and is comparable to hand-written code for most examples.

***Transactional memoory.*** CCRs have been recently reified in several language designs, such as that of Harris and Frasier [24]. Such designs can be viewed as special cases of *transactional memory (TM)* [25], which has attracted enormous attention in the literature, as an alternative of lock-based synchronization [38].

Our implicit-signal monitors are both *less* and *more* ambitious than transactional memory techniques, in different respects. Monitors do not attempt to automate-away lock-based synchronization: we explicitly require the programmer to declare different resources together with their synchronization policy. Concretely, each monitor can be thought of as a single lock, whereas transactional memory techniques do away with distinctions between locks in favor of a single `atomic` construct. The same essential feature is kept in Harris and Frasier's conditional critical regions [24]: although `atomic` statements can have a condition associated with them, their code blocks are guaranteed to execute with atomic semantics, relative to any other `atomic` block, under whichever condition. Therefore, implicit-signal monitors are inherently lower-level than TM approaches, requiring more care on behalf of the programmer, but also imposing no overhead for maintaining atomic semantics. At the same time, our implicit-signal monitors are more ambitious on the implementation side. Static reasoning over abstract conditions results in the removal of extraneous signaling overheads, enabling high-performance execution.

***Automation of synchronization.*** Static analysis has been applied to the optimization of synchronization primitives in the past, mostly in the context of auto-locking or lock-inference techniques [11, 20, 26, 34, 41]. The language model these techniques implement is much closer to transactional memory than implicit-signal monitors: the analysis attempts to infer which locks protect which shared data, as well as which data are not thread-shared. Furthermore, these static analyses typically produce a whole-program model of shared data and do not reason over symbolic conditions, as in our approach. Techniques that infer synchronization given specifications and abstractions [10, 19, 47] often use similar reasoning techniques as our approach (e.g., SMT solvers), but start from much more abstract input and place an emphasis on correctness, not on approaching hand-written code performance. Similar comments apply to approaches that synthesize synchronization actions, given abstract models of program behavior, using control theory techniques [35, 48].

***Program analysis for concurrency.*** A major application of static analysis in the domain of concurrency has been in guaranteeing safety or finding bugs. There are techniques for ensuring safe programming using advanced typing [5], analyses for static race detection [43], approaches to concurrency bug fixing [31], tools that flag suspicious concurrency patterns [29, 49], and much more. Additionally, dynamic techniques for concurrency bug detection [1, 13, 33, 40, 45] often benefit from symbolic reasoning, especially in approaches inspired by model checking or symbolic execution techniques [32, 36, 42]. Such past work is only superficially related to ours, since the aims, programming abstractions, and analysis techniques used are quite different.

***Abductive reasoning in program analysis.*** Our proposed approach uses abductive reasoning for automatically inferring monitor invariants. The use of abduction in program analysis is quite common, especially in the context of modular analysis [2, 9, 18], loop invariant generation [17], and specification inference [2, 50]. However, our use of abductive reasoning differs from prior work in that we use abduction to generate candidate predicates over which we synthesize monitor invariants using predicate abstraction.

## 9 Limitations

While this paper takes a first step towards synthesizing efficient explicit-signal implementations of implicit-signal monitors, it makes several assumptions that allow the proposed approach to be practical. First, we assume that predicates appearing in `waituntil` statements are expressible in some first-order theory (e.g., linear arithmetic). Hence, if the predicate corresponds to the result of a complicated function containing loops or recursion, our approach must be conservative and potentially broadcast at the end of every CCR. However, we believe that such cases are are in practice, allowing Expresso to generate efficient code for monitors taken from real-world applications.

Second, our approach assumes that monitors do not contain nested `waituntil` statements and that calls that occur inside the body of a monitor method are *closed calls* [3], meaning that they do not release access to the monitor. We believe that the latter assumption is quite realistic, since most modern implementations of monitor-like constructs also assume closed-call semantics. On the other hand, handling nested `waituntil` statements is an interesting direction for future work. In particular, we plan to investigate how to leverage more sophisticated forms of symbolic reasoning to analyze implicit-signal monitors with nested `waituntil` statements.

## 10 Conclusion

We have presented a new technique for automatic signalling in concurrent programs. Our method statically analyzes an implicit-signal monitor implementation to determine where signaling is necessary and automatically generates its corresponding explicit-signal version. Our method employs symbolic reasoning to avoid unnecessary context switches and run-time evaluation of predicates. We have implemented the proposed algorithm in a tool called Expresso and evaluate it on 14 benchmarks. Our evaluation shows that the code generated by Expresso is very similar to hand-written code and that it significantly outperforms a state-of-the-art run-time solution in most cases.

## Acknowledgments

We would like to thank Doug Lea, the anonymous reviewers, and the members of the UToPiA group for their insightful feedback. The first three authors of this paper are supported in part by NSF Award #1712060, NSF Award #1453386, AFRL Award #8750-15-2-0096, and the last author is supported by ERC grant 307334 (SPADE), a Facebook Research and Academic Relations Program award, and an Oracle Labs collaborative research grant.

# Appendix A: Well-formedness of Monitor Trace

In this section, we formalize what it means for a monitor trace $\tau$ to be syntactically well-formed. Towards this goal, we first define the projection of a trace $\tau$ onto a thread $t$, denoted as $\tau \downarrow t$.

**Definition 10.1.** The projection of trace $\tau$ onto thread $t$, denoted $\tau \downarrow t$, is defined as follows:

$$
\begin{array}{rcll}
(t', w, true) \downarrow t & = & [w] & \text{if } t' = t \\
(t', w, b) \downarrow t & = & nil & \text{if } t' \neq t \text{ or } b = false \\
e :: \tau \downarrow t & = & (e \downarrow t) :: (\tau \downarrow t) &
\end{array}
$$

**Definition 10.2.** We say that $\tau \downarrow t$ is well-formed if $\tau \downarrow t = [S_1, \ldots, S_n, S_{n+1}]$, each $S_i$ where $i \in [1, n]$ corresponds to the body of some method, and $S_{n+1}$ is a prefix of the body of some method.

Using these definitions, we can now define syntactic well-formedness of monitor traces as follows:

**Definition 10.3.** We say that a monitor trace $\tau$ is syntactically well-formed if the following conditions are satisfied:

1. For every thread $t$, $\tau \downarrow t = S_1; \ldots; S_n; S_{n+1}$ is well-formed
2. If $\tau[i] = (t, w, true)$ and $\tau[i+1] = (t', w', b)$, then either $w$ is the last statement in its method or $t' = t$ and $w'$ is the successor of $w$.

# Appendix B: Soundness Proof

In this section, we prove the soundness of the PlaceSignal algorithm. The proof of Theorem 4.1 follows immediately from Lemmas 10.4 and 10.5.

In the remainder of this section, we use the term *event* to also refer to pairs $(t, w)$ without the corresponding boolean. We refer to $t$ as *thread*($e$) and, if $w = (p, s)$, we refer to $p$ as *pred*($e$). Given two sets $\mathcal{N}, \mathcal{N}'$, we will assume that the total order relation $\prec$ on events is defined in a way that respects the property $\mathcal{N}' \subseteq \mathcal{N} \Rightarrow \min(\mathcal{N}) = \min(\mathcal{N}')$.

Furthermore, for the rest of the section it is useful to keep in mind that the source language maintains the following invariant:

**Invariant** If $(\sigma_0, \tau, \emptyset, \emptyset) \longrightarrow^* (\sigma, \_, \mathcal{B}, \mathcal{N})$, then:
- If $(t, w) \in \mathcal{N}$, then we have $(t, w) \in \mathcal{B}$.
- If $(t, w) \in \mathcal{B}$ and $(\sigma, t) \models Guard(w)$, then $(t, w) \in \mathcal{N}$.

**Lemma 10.4.** Let $M' = PlaceSignals(M, I)$. If $I$ is a correct monitor invariant, then for all monitor states $\sigma$ and all well-formed traces $\tau$, if $(\sigma, \tau, \emptyset, \emptyset) \Longrightarrow^* (\sigma', \epsilon, \_, \_)$ then $(\sigma, \tau, \emptyset, \emptyset) \longrightarrow^* (\sigma', \epsilon, \_, \_)$.

*Proof.* Follows from Lemma 10.8. □

**Lemma 10.5.** Let $M' = PlaceSignals(M, I)$. If $I$ is a correct monitor invariant, then for all monitor states $\sigma$ and all well-formed traces $\tau$. If $(\sigma, \tau, \emptyset, \emptyset) \longrightarrow^* (\sigma', \epsilon, \_, \_)$ and $\tau$ is normalized with respect to $\sigma$, then $(\sigma, \tau, \emptyset, \emptyset) \Longrightarrow^* (\sigma', \epsilon, \_, \_)$.

*Proof.* Follows from Lemma 10.9. □

**Definition 10.6. (Invalidation).** Let $e = (w, t)$ be a monitor event where $w = (p, s)$, $t$ be a thread identifier, and $p$ a predicate. We write $Invalidate(e, t', p')$, if, for any state $\sigma$ such that $\langle s, t, \sigma \rangle \Downarrow \sigma'$, we have $(\sigma', t') \not\models p'$.

**Definition 10.7. (Agreement).** We say that $(\sigma, \mathcal{B}, \mathcal{N})$ agrees with $(\sigma, \mathcal{B}', \mathcal{N}')$, written $(\sigma, \mathcal{B}, \mathcal{N}) \sim (\sigma, \mathcal{B}', \mathcal{N}')'$ if the following conditions are satisfied:

1. $\mathcal{B} = \mathcal{B}'$
2. $\mathcal{N} \supseteq \mathcal{N}'$
3. If $e \in (\mathcal{N} - \mathcal{N}')$, then either
   a. $(\sigma, thread(e)) \not\models pred(e)$, or
   b. $\exists e' \in \mathcal{N}'. e' \prec e \land pred(e) = pred(e') \land Invalidate(e', thread(e), pred(e))$

**Lemma 10.8.** If $(\sigma, \mathcal{B}, \mathcal{N}) \sim (\sigma, \mathcal{B}', \mathcal{N}')$ and
$$(\sigma, \tau, \mathcal{B}', \mathcal{N}') \Longrightarrow (\sigma', \tau', \mathcal{B}_1, \mathcal{N}_1)$$
then we have
$$(\sigma, \tau, \mathcal{B}, \mathcal{N}) \longrightarrow (\sigma', \tau', \mathcal{B}_2, \mathcal{N}_2)$$
and $(\sigma', \mathcal{B}_1, \mathcal{N}_1) \sim (\sigma', \mathcal{B}_2, \mathcal{N}_2)$.

*Proof.* The proof is by induction on $\tau$.

- **Base case 1a:** $\tau = e = (t, w, false)$ and $(t, w) \notin \mathcal{B}'$. In this case, we have $(\sigma, t) \not\models Guard(w)$ and $(\sigma, e, \mathcal{B}', \mathcal{N}') \Longrightarrow (\sigma, \epsilon, \mathcal{B}' \cup \{\overline{e}\}, \mathcal{N}')$. Since $\mathcal{B} = \mathcal{B}'$, we have $(t, w) \notin \mathcal{B}$. Thus, we also have $(\sigma, e, \mathcal{B}', \mathcal{N}) \longrightarrow (\sigma, \epsilon, \mathcal{B}' \cup \{\overline{e}\}, \mathcal{N})$. Since $\mathcal{N}, \mathcal{N}'$ satisfy conditions (2) and (3) of the agreement definition and the state $\sigma$ has not changed, $\mathcal{N}, \mathcal{N}'$ continue to satisfy conditions (2) and (3). Thus, we have $(\sigma, \mathcal{B}' \cup \{\overline{e}\}, \mathcal{N}') \sim (\sigma, \mathcal{B} \cup \{\overline{e}\}, \mathcal{N})$.
- **Base case 1b:** $\tau = e = (t, w, false)$ and $(t, w) \in \mathcal{N}'$. In this case, we have $(\sigma, e, \mathcal{B}', \mathcal{N}') \Longrightarrow (\sigma, \epsilon, \mathcal{B}', \mathcal{N}'\backslash\{\overline{e}\})$. Since $\mathcal{N} \supseteq \mathcal{N}', (t, w) \in \mathcal{N}$, we also have $(\sigma, e, \mathcal{B}', \mathcal{N}) \longrightarrow (\sigma, \epsilon, \mathcal{B}', \mathcal{N}\backslash\{\overline{e}\})$. Since $\mathcal{N}, \mathcal{N}'$ satisfy conditions (2) and (3) of the agreement definition and the state $\sigma$ has not changed, $\mathcal{N}, \mathcal{N}'$ continue to satisfy conditions (2) and (3). Thus, $(\sigma, \mathcal{B}', \mathcal{N}'\backslash\{\overline{e}\}) \sim (\sigma, \mathcal{B}', \mathcal{N}\backslash\{\overline{e}\})$.

- **Base case 2a:** $\tau = e = (t, w, true)$ and $\overline{e} \notin \mathcal{B}'$. In this case, we have:

$$\frac{\begin{array}{c} e = (t, w, true) \\ \overline{e} \notin \mathcal{B}' \quad (\sigma, t) \models Guard(w) \\ \langle Body(w), t, \sigma \rangle \Downarrow \sigma' \\ \mathcal{N}_2 = GetSignals(w, \sigma', \mathcal{B}) \\ \mathcal{N}_3 = GetBroadcasts(w, \sigma', \mathcal{B}) \end{array}}{(\sigma, e, \mathcal{B}', \mathcal{N}') \Longrightarrow (\sigma', \epsilon, \mathcal{B}', \mathcal{N}' \cup \mathcal{N}_2 \cup \mathcal{N}_3)}$$

Since $\mathcal{B} = \mathcal{B}'$, this implies $\overline{e} \notin \mathcal{B}$, thus, we have:

$$\frac{\begin{array}{c} e = (t, w, true) \\ \overline{e} \notin \mathcal{B} \quad (\sigma, t) \models Guard(w) \\ \langle Body(w), t, \sigma \rangle \Downarrow \sigma' \\ \mathcal{N}_1 = \{(t, w) \mid (t, w) \in \mathcal{B}, (\sigma', t) \models Guard(w)\} \end{array}}{(\sigma, e, \mathcal{B}, \mathcal{N}) \longrightarrow (\sigma', \epsilon, \mathcal{B}, \mathcal{N} \cup \mathcal{N}_1)}$$

Part (a) of the agreement definition trivially holds, since $\mathcal{B} = \mathcal{B}'$. For part (b), we need to show that $\mathcal{N}_1 \cup \mathcal{N} \supseteq \mathcal{N}_2 \cup \mathcal{N}_3 \cup \mathcal{N}'$. That is, if $e^* \in \mathcal{N}_2 \cup \mathcal{N}_3 \cup \mathcal{N}'$, we also have $e^* \in \mathcal{N}_1 \cup \mathcal{N}$. Case 1: Suppose $e^* \in \mathcal{N}'$. By agreement, we have $e^* \in \mathcal{N}$, thus part (b) of agreement holds in this case. Case 2: Suppose $e^* \in \mathcal{N}_2 \cup \mathcal{N}_3$. This means $(pred(e^*), c) \in P_1 \cup P_2$ where $P_1 = Signals(w)$ and $P_2 = Broadcasts(w)$. Case 2a: If $c = ?$, then $e^*$ is only added to $\mathcal{N}_2 \cup \mathcal{N}_3$ if $(\sigma', thread(e^*)) \models pred(e^*)$. But then, we have $e^* \in \mathcal{N}_1$. Case 2b: If $c = \checkmark$, then the PlaceSignals algorithm ensures that $\vdash \{I \land \neg pred(e^*)\} Body(w) \{pred(e^*)\}$. By correctness of this Hoare triple, we have $(\sigma', thread(e^*)) \models pred(e^*)$. Thus, $e^* \in \mathcal{N}_1$.

For part (c) of the agreement definition, we need to show that if $e^* \in (\mathcal{N} \cup \mathcal{N}_1) \backslash (\mathcal{N}' \cup \mathcal{N}_2 \cup \mathcal{N}_3)$, then either (i) $(\sigma', thread(e^*)) \not\models pred(e^*)$, or (ii) $\exists e' \in \mathcal{N}_2 \cup \mathcal{N}_3 \cup \mathcal{N}$ such that $e' \prec e^* \land pred(e) = pred(e^*) \land Invalidate(e', thread(e^*), pred(e^*))$. We consider two cases:

– Case 1: $e^* \in \mathcal{N}$, but $e^* \notin \mathcal{N}' \cup \mathcal{N}_2 \cup \mathcal{N}_3$ (i.e., $e^* \notin \mathcal{N}'$, $e^* \notin \mathcal{N}_2$, $e^* \notin \mathcal{N}_3$). Since $e^* \in \mathcal{N} - \mathcal{N}'$, we have (from agreement) that either (a) $(\sigma, thread(e^*)) \not\models pred(e^*)$, or
(b) $\exists e' \in \mathcal{N}'$ such that $e' \prec e^*, pred(e') = pred(e^*)$ and $Invalidate(e', thread(e^*), pred(e^*))$.
  * Case 1a: $(\sigma, thread(e^*)) \not\models pred(e^*)$. If we also have $(\sigma', thread(e^*)) \not\models pred(e^*)$, then part (c) of agreement definition obviously holds. If $(\sigma', thread(e^*)) \models pred(e^*)$, then $e^* \in \mathcal{N}_1$, and we argue correctness in Case 2.
  * Case 1b: $(\sigma, thread(e^*)) \not\models pred(e^*)$ and $\exists e' \in \mathcal{N}'$ such that $(e' \prec e^*, pred(e') = pred(e^*)$, and $Invalidate(e', thread(e^*), pred(e^*))$. Since such an $e'$ is also in $\mathcal{N}' \cup \mathcal{N}_2 \cup \mathcal{N}_3$, part (c) of the agreement definition is preserved.
– Case 2: $e^* \in \mathcal{N}_1$, but $e^* \notin \mathcal{N}'$, $e^* \notin \mathcal{N}_2$, and $e^* \notin \mathcal{N}_3$. There are only two ways in which $e^* \in \mathcal{N}_1$ but not in $\mathcal{N}_2$ or $\mathcal{N}_3$:
  * Case 2a: $(pred(e^*), c) \notin P_1 \cup P_2$ where $P_1 = Signals(w)$ and $P_2 \in Broadcasts(w)$. Using the invariant of the PlaceSignals algorithm, this can only happen if $\vdash \{I \land \neg pred(e^*)\} Body(w) \{\neg pred(e^*)\}$. If $pred(e^*)$ was

true before executing $Body(w)$ (i.e., $(\sigma, thread(e^*)) \models pred(e^*)$), then $e^*$ would also be in $\mathcal{N}$, which implies it is also in $\mathcal{N}'$, so this case is not possible. Otherwise, if $(\sigma, thread(e^*)) \not\models pred(e^*)$, the correctness of the Hoare triple implies that $(\sigma', thread(e^*)) \not\models pred(e^*)$, which contradicts the fact that $e^* \in \mathcal{N}_1$.

  * Case 2b: $(pred(e^*), c) \in P_2$, but there exists another $e'$ such that (i) $e' \prec e^*$, and (ii) $pred(e^*) = pred(e)$. The only way in which $e^*$ is added to $P_2$ but not $P_1$ by the PlaceSignals algorithm is if the following Hoare triple holds for all waituntil statements $w$ with guard $pred(e^*)$:

    $\{I \wedge Guard(w)\}\ Body(w)\ \{\neg Guard(w)\}\}$

    Since $e'$ has the same guard as $pred(e^*)$, the validity of the Hoare triple implies $Invalidate(e', thread(e^*), pred(e^*))$. Thus part (c) of the agreement definition again holds.

- **Base case 2b:** $\tau = e = (t, w, true)$ and $\bar{e} = min(\mathcal{N}')$. In this case, we have:

$$\frac{\begin{array}{c} e = (t, w, true) \\ \bar{e} = min(\mathcal{N}') \quad (\sigma, t) \models Guard(w) \\ \langle Body(w), t, \sigma \rangle \Downarrow \sigma' \\ \mathcal{N}_2 = GetSignals(w, \sigma', \mathcal{B}) \\ \mathcal{N}_3 = GetBroadcasts(w, \sigma', \mathcal{B}) \end{array}}{(\sigma, e, \mathcal{B}, \mathcal{N}') \Longrightarrow (\sigma', \epsilon, \mathcal{B} \backslash \{\bar{e}\}, (\mathcal{N}' \cup \mathcal{N}_2 \cup \mathcal{N}_3) \backslash \{\bar{e}\})}$$

Since $\mathcal{N}' \subseteq \mathcal{N}$, we have $min(\mathcal{N}) = min(\mathcal{N}')$; hence:

$$\frac{\begin{array}{c} e = (t, w, true) \\ \bar{e} = min(\mathcal{N}) \quad (\sigma, t) \models Guard(w) \\ \langle Body(w), t, \sigma \rangle \Downarrow \sigma' \\ \mathcal{N}_1 = \{(t, w) \mid (t, w) \in \mathcal{B}, (\sigma', t) \models Guard(w)\} \end{array}}{(\sigma, e, \mathcal{B}, \mathcal{N}) \longrightarrow (\sigma', \epsilon, \mathcal{B} \backslash \{\bar{e}\}, (\mathcal{N} \cup \mathcal{N}_1) \backslash \{\bar{e}\})}$$

Because the sets $\mathcal{N}_1, \mathcal{N}_2, \mathcal{N}_3$ are constructed in exactly the same way as Base case 2a, the same argument also applies in this case. However, we additionally need to show that if $e^* \in \mathcal{N} \backslash \mathcal{N}'$, and $pred(e) = pred(e^*)$, then $(\sigma', thread(e^*)) \not\models pred(e^*)$. Since $e \in \mathcal{N}'$ and $pred(e^*) = pred(e)$ and $e \prec e^*$, we have $Invalidate(e, thread(e^*), pred(e))$. By definition of the $Invalidate$ predicate, this means that $(\sigma', thread(e^*)) \not\models pred(e^*)$.

- **Inductive step:** If trace $\tau$ contains multiple events, then we have:

$$\frac{(\sigma', e, \mathcal{B}', \mathcal{N}') \Longrightarrow (\sigma_2, \epsilon, \mathcal{B}_2, \mathcal{N}_2)}{(\sigma', e :: \tau, \mathcal{B}', \mathcal{N}') \Longrightarrow (\sigma_2, \tau, \mathcal{B}_2, \mathcal{N}_2)}$$

and

$$\frac{(\sigma, e, \mathcal{B}, \mathcal{N}) \longrightarrow (\sigma_1, \epsilon, \mathcal{B}_1, \mathcal{N}_1)}{(\sigma, e :: \tau, \mathcal{B}, \mathcal{N}) \longrightarrow (\sigma_1, \tau, \mathcal{B}_1, \mathcal{N}_1)}$$

Using the inductive hypothesis and the assumption that $(\sigma', \mathcal{B}', \mathcal{N}') \sim (\sigma, \mathcal{B}, \mathcal{N})$, we have $(\sigma_2, \mathcal{B}_2, \mathcal{N}_2) \sim (\sigma_1, \mathcal{B}_1, \mathcal{N}_1)$.

□

**Lemma 10.9.** *If* $(\sigma, \mathcal{B}, \mathcal{N}) \sim (\sigma, \mathcal{B}', \mathcal{N}')$, $\tau$ *is normalized, and*

$$(\sigma, \tau, \mathcal{B}, \mathcal{N}) \longrightarrow (\sigma', \tau', \mathcal{B}_1, \mathcal{N}_1)$$

*then we have*

$$(\sigma, \tau, \mathcal{B}', \mathcal{N}') \Longrightarrow (\sigma', \tau', \mathcal{B}_2, \mathcal{N}_2)$$

*and* $(\sigma', \mathcal{B}_1, \mathcal{N}_1) \sim (\sigma', \mathcal{B}_2, \mathcal{N}_2)$.

*Proof.* The proof is by induction on $\tau$.

- **Base case 1a:** $\tau = e = (t, w, false)$ and $(t, w) \notin \mathcal{B}$. In this case, we have $(\sigma, t) \not\models Guard(w)$ and $(\sigma, e, \mathcal{B}', \mathcal{N}') \longrightarrow (\sigma, \epsilon, \mathcal{B}' \cup \{\bar{e}\}, \mathcal{N}')$. Since $\mathcal{B} = \mathcal{B}'$, we have $(t, w) \notin \mathcal{B}'$. Thus, we also have $(\sigma, e, \mathcal{B}', \mathcal{N}) \Longrightarrow (\sigma, \epsilon, \mathcal{B}' \cup \{\bar{e}\}, \mathcal{N})$. Since $\mathcal{N}, \mathcal{N}'$ satisfy conditions (2) and (3) of the agreement definition and the state $\sigma$ has not changed, $\mathcal{N}, \mathcal{N}'$ continue to satisfy conditions (2) and (3). Thus, we have $(\sigma, \mathcal{B}' \cup \{\bar{e}\}, \mathcal{N}') \sim (\sigma, \mathcal{B} \cup \{\bar{e}\}, \mathcal{N})$.

- **Base case 1b:** $\tau = e = (t, w, false)$ and $(t, w) \in \mathcal{N}$. We do not need to consider this case because it contradicts the assumption that the trace is normalized.

- **Base case 2a:** $\tau = e = (t, w, true)$ and $\bar{e} \notin \mathcal{B}$. In this case, we have:

$$\frac{\begin{array}{c} e = (t, w, true) \\ \bar{e} \notin \mathcal{B} \quad (\sigma, t) \models Guard(w) \\ \langle Body(w), t, \sigma \rangle \Downarrow \sigma' \\ \mathcal{N}_1 = \{(t, w) \mid (t, w) \in \mathcal{B}, (\sigma', t) \models Guard(w)\} \end{array}}{(\sigma, e, \mathcal{B}, \mathcal{N}) \longrightarrow (\sigma', \epsilon, \mathcal{B}, \mathcal{N} \cup \mathcal{N}_1)}$$

Since $\mathcal{B} = \mathcal{B}'$, this implies $\bar{e} \notin \mathcal{B}'$, thus, we also have:

$$\frac{\begin{array}{c} e = (t, w, true) \\ \bar{e} \notin \mathcal{B}' \quad (\sigma, t) \models Guard(w) \\ \langle Body(w), t, \sigma \rangle \Downarrow \sigma' \\ \mathcal{N}_2 = GetSignals(w, \sigma', \mathcal{B}) \\ \mathcal{N}_3 = GetBroadcasts(w, \sigma', \mathcal{B}) \end{array}}{(\sigma, e, \mathcal{B}', \mathcal{N}') \Longrightarrow (\sigma', \epsilon, \mathcal{B}', \mathcal{N}' \cup \mathcal{N}_2 \cup \mathcal{N}_3)}$$

Thus, the proof is exactly the same as Base Case 2a of Lemma 10.8

- **Base case 2b:** $\tau = e = (t, w, true)$ and $\bar{e} = min(\mathcal{N})$. In this case, we have:

$$\frac{\begin{array}{c} e = (t, w, true) \\ \bar{e} = min(\mathcal{N}) \quad (\sigma, t) \models Guard(w) \\ \langle Body(w), t, \sigma \rangle \Downarrow \sigma' \\ \mathcal{N}_1 = \{(t, w) \mid (t, w) \in \mathcal{B}, (\sigma', t) \models Guard(w)\} \end{array}}{(\sigma, e, \mathcal{B}, \mathcal{N}) \longrightarrow (\sigma', \epsilon, \mathcal{B} \backslash \{\bar{e}\}, (\mathcal{N} \cup \mathcal{N}_1) \backslash \{\bar{e}\})}$$

The existence of $e$ in $\mathcal{N}$ does not guarantee the existence of $e$ in $\mathcal{N}'$. We consider two cases, namely (1) $e \in \mathcal{N}'$ and (2) $e \notin \mathcal{N}'$. For case (1), the explicit-signal transitions also use rule (2b), thus the proof is exactly the same as Case 2b of the proof of Lemma 10.8.

We will now argue that case 2 (i.e., $e \notin \mathcal{N}'$) is not possible. Suppose $e \in \mathcal{N}$, but not in $\mathcal{N}'$. Then, by part (c) of the agreement definition, we have either (i) $(\sigma, t) \not\models Guard(w)$, or $\exists e' \in \mathcal{N}'. \ e' \prec e \wedge pred(e) = pred(e') \wedge Invalidate(e', t, pred(e))$. Observe that (i) contradicts the assumption $(\sigma, t) \models Guard(w)$. For (ii), we have $\exists e' \in \mathcal{N}'. \ e' \prec e \wedge pred(e) = pred(e') \wedge Invalidate(e', t, pred(e))$. But since $\mathcal{N} \supseteq \mathcal{N}'$, $e'$ is also in $\mathcal{N}$ and furthermore $e' \prec e$. But this contradicts the assumption that $e'$ is the minimum element of $\mathcal{N}$.
- **Inductive step:** Same as the inductive step of the proof of Lemma 10.8.

$\square$

## Appendix C: Proof of Theorem 4.3

The proof is by induction on the length of $\tau_0$. For the base case, we have $\tau_0 = \epsilon$, thus $\tau' = \tau$ and the statement trivially holds. For the inductive step, we have $\tau_0 = e'\tau_1$. Suppose $(\sigma, e'\tau_1 e, \mathcal{B}, \mathcal{N}) \longrightarrow^* (\sigma, \epsilon, \mathcal{B}', \mathcal{N}')$ where $(\sigma, e', \mathcal{B}, \mathcal{N}) \longrightarrow^* (\sigma_1, \epsilon, \mathcal{B}_1, \mathcal{N}_1)$ as well as $(\sigma_1, \tau_1 e, \mathcal{B}_1, \mathcal{N}_1) \longrightarrow^* (\sigma', \epsilon, \mathcal{B}', \mathcal{N}')$. By the inductive hypothesis, we have $(\sigma_1, e\tau_1, \mathcal{B}_1, \mathcal{N}_1) \longrightarrow^* (\sigma', \epsilon, \mathcal{B}', \mathcal{N}')$. This implies $(\sigma, e'e\tau_1, \mathcal{B}, \mathcal{N}) \longrightarrow^* (\sigma', \epsilon, \mathcal{B}', \mathcal{N}')$. Finally, from the assumption $Comm(Body(e), M)$, we have $Body(e); Body(e') \equiv Body(e'); Body(e)$. Thus, we also have $(\sigma, ee'\tau_1, \mathcal{B}, \mathcal{N}) \longrightarrow^* (\sigma', \epsilon, \mathcal{B}', \mathcal{N}')$

## Appendix D: Sample Monitor Invariant

To give the reader some idea about the inferred monitor invariants, we show the invariant (in SMTLIB format) for the AsyncDispatch benchmark from Gradle:

```
(let ((a!1 (not (= (queue.size)
                    0)))
      (a!3 (not (>= (queue.size))
                    (+ 1
                       (maxQueueSize)))))
      (a!5 (not (>= (queue.size))
                    (maxQueueSize)))))
(let ((a!2 (not (or (= (state)
                        Stopped)
                    a!1)))
      (a!4 (or (= (queue.size)
                  0)
               (= (queue.first)
                  0)
               (= (state)
                  Stopped)
               a!3)))
(let ((a!6 (and a!4
                (or a!1
                    (= (state)
                        Stopped)
                    a!5))))
  (and (or a!2 a!6)
       (not (<= (maxQueueSize)
                0))
       (>= (maxQueueSize)
           0)
       (>= (queue.size)
           0)))))
```

As another example, we show the monitor invariant for the BoundedBuffer benchmark:

```
(let ((a!1 (not (>= buff.length
                     0)))
      (a!2 (not (>= (count)
                     buff.length)))
      (a!4 (>= (count)
               (+ 1
                  (buff.length)))))
(let ((a!3 (or a!1
               (not (or a!1 a!2))
               (not (<= (count)
                        (- 1))))))
  (and a!3
       (>= (count)
           0)
       (or a!1
           (<= (count)
               0)
           (not a!4)))))
```